\documentclass[twocolumn,aps,superscriptaddress,showpacs]{revtex4}

\usepackage{amssymb}
\usepackage{amsmath}
\usepackage{graphicx}
\usepackage[normalem]{ulem}
\usepackage[dvips]{color}
\usepackage{multirow}
\usepackage{appendix}

\makeatletter

\newcommand{\Rmnum}[1]{\expandafter\@slowromancap\romannumeral #1@}
\makeatother

\makeatletter 
\@addtoreset{equation}{section}
\makeatother  

\begin{document}

\title{Investigating the scaling of higher-order flows in relativistic heavy-ion collisions}
\author{Chun-Jian Zhang}
\affiliation{Shanghai Institute of Applied Physics, Chinese Academy
of Sciences, Shanghai 201800, China}
\affiliation{University of Chinese Academy of Sciences, Beijing 100049, China}
\author{Jun Xu\footnote{corresponding author: xujun@sinap.ac.cn}}
\affiliation{Shanghai Institute of Applied Physics, Chinese Academy
of Sciences, Shanghai 201800, China}
\date{\today}

\begin{abstract}

The modified number of constituent quark (NCQ) scaling
$v_{n}/n_{q}^{n/2} \sim KE_{T}/n_{q}$ for mesons and baryons and the
scaling relation $v_{n} \sim v_{2}^{n/2}$ for higher-order
anisotropic flows, which were observed experimentally, have been
investigated at top RHIC energy. It has been found that the modified
NCQ scaling can not be obtained from the naive coalescence even by
taking into account event-by-event fluctuations but may be due to
hadronic afterburner or thermal freeze-out. In addition, we observed
that the behavior of the $v_{n}/v_{2}^{n/2}$ ratio is sensitive to
the partonic interaction, and it is
different for mesons and baryons from the naive coalescence but is expected to be almost the same experimentally.

\end{abstract}


\maketitle
\section{Introduction}
\label{introduction}

Relativistic heavy-ion collisions provide a useful way of studying
the new phase which may exist at extreme high energy densities on
earth. Collectivity is one of the main evidences of the produced
dense matter, named as quark-gluon plasma (QGP), produced in
relativistic heavy-ion collider
(RHIC)~\cite{Ars05,Bac05,Ada05a,Adc05}. Due to the almond shape of
the produced QGP in non-central collisions, there are more
freeze-out particles moving in-plane than out-plane, leading to the
so-called elliptic flow ($v_2$). Generally, the anisotropic flow is
believed to be mostly produced at the early stages of the collision
when the interaction is strongest. The number of constituent quark
(NCQ) scaling law $v_2/n_q \sim p_T/n_q$ or $v_2/n_q \sim
KE_T/n_q$~\cite{STAR04,STAR05a,STAR05b,PHENIX07a,PHENIX07b,STAR08},
where $p_T$ and $KE_T$ are respectively the transverse momentum and
transverse kinetic energy, shows that the underlining mechanism for
hadron elliptic flow is from partons as baryons and mesons are
scaled by their number of constituent quark numbers $n_q$. The NCQ
scaling can be well explained by the coalescence
model~\cite{Gre03a,Gre03b,Fri03a,Fri03b,Hwa03}, typically by
assuming that hadrons are formed from the combination of its
constituent quarks whose distance in momentum space is
small~\cite{Mol03,Kol04}. The coalescence or recombination mechanism
also automatically results in the relation $v_4 \sim v_2^2$ from the
leading order~\cite{Kol04}, which originates actually from the
partonic level~\cite{Che04}. Although the above coalescence picture
works well at intermediate $p_T$ or $KE_T$, it has been observed
that the collective flows of light and heavy hadrons obey the mass
ordering at low $p_T$ showing the thermalization of different
species of particles in the
medium~\cite{STAR05a,PHENIX07a,PHENIX07b}, and this can usually be
explained by a blast wave model~\cite{Lis04} or the Cooper-Frye
freeze-out condition~\cite{Coo74} in the hydrodynamic model.

Recently, it was realized that the initial spatial distribution of
QGP is not a smooth one but has density
fluctuations~\cite{Alv10a,Alv10b}. The initial anisotropy in
coordinate space can develop into the final anisotropy in momentum
space as a result of QGP interaction. This leads to the redefinition
of higher-order harmonic flows with respect to their event plane or
participant plane, especially the odd-order harmonic
flows~\cite{Alv10a,Alv10b,Pet10,Sch11,Xu11a,Xu11b,Ma11}. It was
further found that the scaling of the higher-order harmonic flows is
modified to $v_n/n_q^{n/2} \sim KE_T/n_q$ mostly from least square
fit~\cite{PHENIX14}. The above modified NCQ scaling for higher-order
flows might stem from the relation between flows of different orders
$v_n \sim v_2^{n/2}$~\cite{ATLAS12}, although the relationship between the two
scalings has never been clarified. It was further found that the
coefficient is nearly a constant with respect to transverse momentum
but increases with decreasing collision centrality~\cite{ATLAS12,Lac11a}. Studying the
scaling relation between flows of different orders is helpful in understand the behavior of the initial eccentricities~\cite{Lac10,Lac11b}, the viscous property of QGP~\cite{Gar12}, and the acoustic nature of anisotropic flows~\cite{Lac11a},
while it is known that the hadronization may affect the scaling
coefficient. Since in the previous recombination model~\cite{Kol04}
the initial density fluctuation was not considered, it is of great
interest to include event plane corrections in the quark coalescence
model. In the present work we carry out such a study to see whether
the corrections can lead to the modified scaling of the
higher-order harmonic flows $v_n/n_q^{n/2} \sim KE_T/n_q$. We also
try to investigate the scaling $v_n \sim v_2^{n/2}$ and understand
the relation between the two scalings. We will see that the modified
scaling relation can originate from the hadronic afterburner or
thermalization in the freeze-out stage instead of higher-order
corrections in the coalescence picture.

This paper is organized as follows. In Sec.~\ref{models} we briefly
describe the models and formalism used in the present study, i.e., a
multiphase transport (AMPT) model, the quark coalescence formalism
with event-by-event fluctuations, and the thermal blast wave model.
In Sec.~\ref{results} we investigate the scaling law of
$v_n/n_q^{n/2} \sim KE_T/n_q$ and $v_n \sim v_2^{n/2}$ in detail by
using the theoretical tool presented in Sec.~\ref{models}. Finally,
a conclusion is given in Sec.~\ref{summary}.

\section{Models and formalism}
\label{models}

In the present study, the AMPT model is used to give a reasonable
final parton phase-space distribution and serves as a useful tool to
test the effect of hadronic afterburner. With the collective flows
of partons in the freeze-out stage, a naive quark coalescence model
is used to generate the hadronic flows analytically in the spirit of
Ref.~\cite{Kol04} by taking into account event-by-event
fluctuations. To study the scaling law from thermal freeze-out other than the coalescence picture, a
generalized blast wave model with higher-order flows is also
described for the convenience of discussion.

\subsection{AMPT Model}

The AMPT model~\cite{Lin05} has been widely used in theoretical
studies or experimental simulations. For Au+Au collisions at
$\sqrt{s_{NN}}=200$ GeV, which is the system in the present study,
the string melting version of AMPT is used. The initial parton
information is generated from the Heavy-Ion Jet INteration Generator
(HIJING) model~\cite{Wan91} by melting hadrons into their valence
quarks and antiquarks. The evolution of the partonic phase is then
modeled by Zhang's Parton Cascade (ZPC) model~\cite{Zha98}, where
the interaction between quarks or antiquarks is effectively
described by two-body scatterings. The freeze-out time of a parton
is given by its last scattering, after which the distance between
two quarks is out of their scattering cross section. The phase-space
information at this stage is used for hadronization in AMPT and the
analytical coalescence as given in the next subsection. In the
current version of AMPT, the hadronization is described by a
coalescence model in which quarks or antiquarks that are close in
coordinate space can form hadrons, and the hadron species depends on
the flavors of its valence quarks and their invariant mass. In this
way the space anisotropy from the final partonic stage to the
initial hadronic stage is preserved, while the distance between
valence quarks in momentum space may not be small. We will return to
this point later. After hadronization, the hadronic evolution is
described by a relativistic transport (ART) model~\cite{Li95}, where
elastic and inelastic scatterings as well as resonance decays of
hadrons are properly treated. We will use ART as a tool to
investigate the hadronic afterburner effect on the modified NCQ
scaling in the present work.

\subsection{Analytical coalescence}

The above described AMPT model is a dynamical transport model. To
have some insights into the quark coalescence mechanism from a more
easily handling way, here we give the analytical coalescence
formalism by extending the previous work in Ref.~\cite{Kol04} and
taking into account event-by-event fluctuations. We start from the
following azimuthal distribution of partons at freeze-out stage:
\begin{eqnarray}
f(p_{T},\phi) \propto 1 + 2\sum_{n=1}^{\infty} v _{n} \cos [n(\phi - \psi_{n})],
\end{eqnarray}%
where $\phi$ is the azimuthal angle, $v_n$ is the $n$th-order
anisotropic flow, and $\psi_{n}$ is the corresponding event plane.
In the present study, the partonic flow $v_n$ can be obtained from the
AMPT model with $\psi_{n}$ determined by the parton phase-space
distribution at the freeze-out stage.

In the naive analytical coalescence picture, the momentum
distribution of quarks inside hadrons is neglected, and the hadron
yield is proportional to the quark density to the power of its
constituent quark number. This can be viewed as a limit in the dynamical coalescence
method~\cite{Gre03a,Gre03b} where the momentum part of the Wigner function is a $\delta$ function instead of a Gaussian form. In this limit
the azimuthal distribution of mesons and baryons can be expressed respectively as
\begin{eqnarray}
F(2p_{T},\phi) &\propto& f^{2}(p_{T},\phi) \propto 1 + 2\sum_{n=1}^{\infty}V_{n} \cos [n(\phi - \psi_{n})], \\
\widetilde{F}(3p_{T},\phi) &\propto& f^{3}(p_{T},\phi) \propto 1 + 2\sum_{n=1}^{\infty}\widetilde{V}_{n} \cos[n(\phi - \psi_{n})],
\end{eqnarray}%
where the anisotropic flows of mesons and baryons can be calculated
respectively from
\begin{eqnarray}
V_{n} = \frac{\int_{0}^{2\pi} \cos(n\phi - n\psi_{n})F(2p_{T},\phi)d\phi}{\int_{0}^{2\pi}F(2p_{T},\phi)d\phi}
\end{eqnarray}%
and
\begin{eqnarray}
\widetilde{V}_{n} &=& \frac{\int_{0}^{2\pi} \cos(n\phi - n\psi_{n})\widetilde{F}(3p_{T},\phi)d\phi}{\int_{0}^{2\pi}\widetilde{F}(3p_{T},\phi)d\phi}.
\end{eqnarray}%
In the above expressions we omit the transverse momentum dependence
of the anisotropic flows, i.e., $v_n(p_T)$, $V_n(2p_T)$, and
$\widetilde{V}_{n}(3p_T$). We expand the value of $n$ up to 4 in
this work. The detailed expressions of meson and baryon flows as well
as their various ratios in terms of the partonic flows and event
planes are given in \ref{app}.

\subsection{Blast Wave Model}
\label{bw}

In the present subsection, we briefly review the blast wave model
which can be viewed as a simplified version of the Cooper-Frye
freeze-out condition used in the hydrodynamic model. In this sense,
the initial hadrons right after hadronization are assumed to be in
thermal and chemical equilibrium, and the medium is undergoing a
collective expansion. In the 'standard' version of the blast wave
model~\cite{Lis04}, particles are emitted perpendicularly from the
surface of an elliptical medium in the transverse plane representing
the azimuthal distribution in the mid-rapidity region, and this
model can be used to fit the $p_T$ spectra and the $v_2$ of
different particle species reasonably well, by neglecting the
hadronic afterburner effect. The 'standard' version of the blast
wave model can be easily generalized to include the higher-order
collective flows and anisotropies of the system right after
hadronization~\cite{PHENIX14}.

In the generalized blast wave model, the lorentz-invariant thermal
distribution can be expressed as
\begin{equation}
f(\vec{r},\vec{p}) \propto \exp(-p^{\mu}u_{\mu}/T_{f}),
\end{equation}
where $T_f$ is the freeze-out temperature, $p^\mu=\{E,p_x,p_y,p_z\}$
is the four-momentum, $u_\mu=\gamma\{1,\rho_x,\rho_y,\rho_z\}$ is
the four-velocity field with
$\gamma=1/\sqrt{1-\rho_x^2-\rho_y^2-\rho_z^2}$, and the $n$th-order
azimuthal velocity as well as the spatial density anisotropies are respectively
expressed as
\begin{eqnarray}
\rho(n,\phi,r) &=& \rho_{0}\{1+ \sum_{n=1}^{\infty}\rho_{n} \cos [n(\phi - \psi_{n})]\}\frac{r}{R}, \\
S(n,\phi) &=& 1+ \sum_{n=1}^{\infty} s_{n} \cos [n(\phi - \psi_{n})].
\end{eqnarray}%
In the above, $\rho_0$ is the radial flow, $R$ is the size of the
emission source, and $\psi_{n}$ is the event plane but is set to 0
in the blast wave study. In the hydrodynamical calculation, both the
velocity and spatial anisotropy coefficients $\rho_n$ and $s_n$ can
be consistently obtained. Since in the present study we will only
consider the modified NCQ scaling $v_n/n_q^{n/2} \sim KE_T/n_q$ for
mesons and baryons using the generalized blast wave model, we simply
set $\rho_n$ and $s_n$ to be the same for different orders $n$. The
values of the parameters are taken from Ref.~\cite{Oh09} used to
describe the initial hadron distribution before hadronic evolution
in Au+Au collisions at $\sqrt{s_{NN}}=200$ GeV, and they are
$T_f=175$ MeV, $R=5.0$ fm, $\rho_{0}=0.55$, $\rho_{n}=0.43$, and
$s_{n}=-0.05$ fm.

\section{Results and discussions}
\label{results}

We now investigate the scaling of higher-order anisotropic flows
$v_n$ in detail. The standard event-plane method in calculating
$v_n$ is detailed in \ref{vn}, where the auto-correlation between
the particle and the event plane is removed, and it is found that
the resolution correction is very small. In our previous
work~\cite{Xu11b}, a partonic scattering cross section of 1.5 mb in
the AMPT model is used to describe the experimental anisotropic
flows from two-particle cumulant method in Au+Au collisions at
$\sqrt{s_{NN}}=200$ GeV. Since the purpose of this study is not to
fit the experimental data but to understand the origin of the
scaling law of higher-order harmonic flows, we will compare the
results from partonic scattering cross sections of 1.5 mb and 10 mb,
and will mainly focus on the results from the cross section of 10 mb
with a large collectivity effect.

\subsection{The modified NCQ scaling $v_{n}/n_{q}^{n/2} \sim KE_{T}/n_{q}$}

\begin{figure}[h]
\includegraphics[scale=0.8]{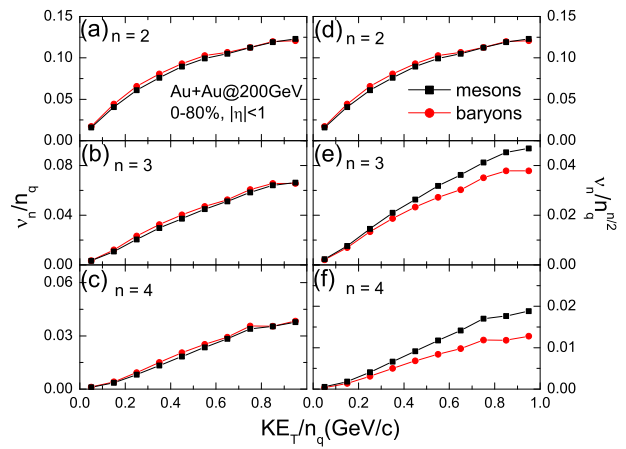} \caption{(Color online)
Scaling relations of hadrons $v_{n} \sim KE_{T}$ in mini-bias Au+Au
collisions at $\sqrt{s_{NN}}=200$ GeV from the analytical
coalescence scenario. } \label{vnket_naive}
\end{figure}

First of all, we investigate whether the higher-order corrections
from event-by-event fluctuations in the analytical coalescence
scenario may lead to the modified NCQ scaling $v_{n}/n_{q}^{n/2}
\sim KE_{T}/n_{q}$. In this case events of Au+Au collisions at
$\sqrt{s_{NN}}=200$ GeV with a partonic scattering cross section of
10 mb have been generated from the AMPT model to get the information
of partonic flows at freeze-out. The hadronic flow is then
calculated through the analytical coalescence scenario. The mass of
the hadron used in calculating the transverse kinetic energy
$KE_{T}=\sqrt{p_{T}^{2}+m^{2}}-m$ is set to be twice or three times
the bare quark mass in ZPC but ideally it should approach the
constituent mass at hadronization, with the latter realized in a
more realistic Nambu-Jona-Lasinio transport model~\cite{Son12,Xu14}.
We note the momentum is conserved while the energy conservation is
violated in the coalescence model. It is found that only the leading
terms in Eqs.~(\ref{v2m}), (\ref{v3m}), (\ref{v4m}), (\ref{v2b}),
(\ref{v3b}), and (\ref{v4b}) are important, while the higher-order
terms are negligibly small which do not account for the modified NCQ
scaling. According to Fig.~\ref{vnket_naive}, it is seen that the analytical
coalescence scenario leads to the original NCQ scaling $v_{n}/n_{q}
\sim KE_{T}/n_{q}$ instead of the modified one $v_{n}/n_{q}^{n/2}
\sim KE_{T}/n_{q}$.

To understand the relation between the modified NCQ scaling
$v_{n}/n_{q}^{n/2} \sim KE_{T}/n_{q}$ and the scaling relation
$v_{n} \sim v_{2}^{n/2}$, we can go into further details of the
results in Fig.~\ref{vnket_naive} in a semi-analytical way. Suppose for
mesons and baryons we have the scaling relation
\begin{eqnarray}
v_{n} &=& C_{n}^{m}v_{2}^{n/2}, \text{(for mesons)} \nonumber\\
v_{n} &=& C_{n}^{b}v_{2}^{n/2}, \text{(for baryons)} \nonumber\\
\end{eqnarray}%
for $n>2$ with the scaling coefficients $C_{n}^{m}$ and $C_{n}^{b}$
for mesons and baryons, respectively. Then, if the NCQ scaling for
$v_2$ is satisfied, i.e., $v_{2}/n_{q} = g(KE_{T}/n_{q})$, we
automatically get the modified NCQ scaling relation for higher-order
flows ($n>2$)
\begin{eqnarray}
v_{n}^{m}/n_{q}^{n/2} &=& C_{n}^{m}g^{n/2}(KE_{T}/n_{q}), \text{(for mesons)} \nonumber\\
v_{n}^{b}/n_{q}^{n/2} &=& C_{n}^{b}g^{n/2}(KE_{T}/n_{q}). \text{(for baryons)}\nonumber\\
\end{eqnarray}%
The modified NCQ scaling relation is satisfied only if
$C_{n}^{m}=C_{n}^{b}$, which is not the case from the analytical
coalescence scenario. If the value of $F_0/\widetilde{F_0}$
is approximated to be 1, $C_{n}^{m}/C_{n}^{b}$ is about $\sqrt{3/2}$ for $n=3$
and $3/2$ for $n=4$, from the leading terms in Eqs.~(\ref{r3m}),
(\ref{r4m}), (\ref{r3b}), and (\ref{r4b}).

\begin{figure}[h]
\includegraphics[scale=0.8]{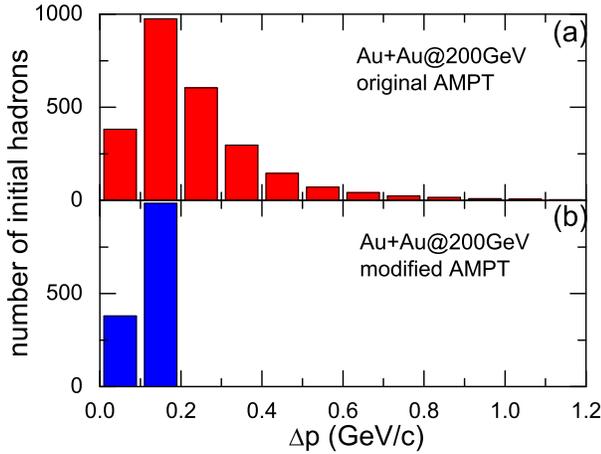} \caption{(Color online)
Histogram of the momentum distance $\Delta p$ between valence quarks
in the hadronization process from the original and the modified AMPT
model in Au+Au collisions at $\sqrt{s_{NN}}=200$ GeV. } \label{histogram}
\end{figure}

\begin{figure}[h]
\includegraphics[scale=0.8]{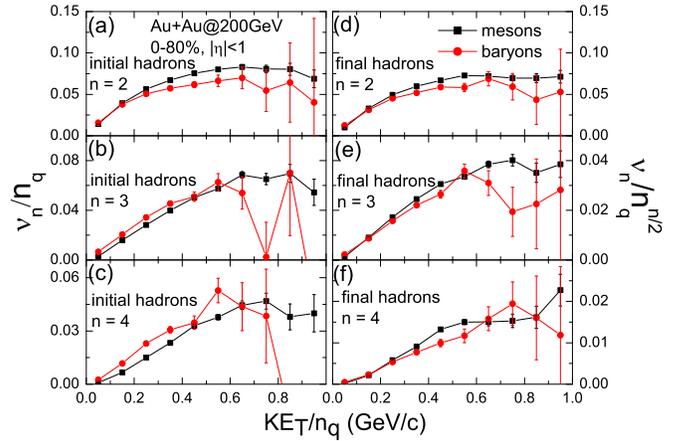} \caption{(Color online) Scaling relation of
$v_{n} \sim KE_{T}$ for initial hadrons right after hadronization
(left) and final hadrons after hadronic evolution (right) in
mini-bias Au+Au collisions at $\sqrt{s_{NN}}=200$ GeV from the
original AMPT model. } \label{vnket_ampt_ori}
\end{figure}

It has been shown that the flows of mesons and baryons from the AMPT
model show a reasonable modified NCQ scaling relation in
Ref.~\cite{Han11}, although the reason has never been clarified. As
we have mentioned, in the original AMPT model partons which are
closer in coordinate space can coalesce into hadrons to preserve
the geometry distribution, while the momentum distance $\Delta p$
between valence quarks may not be small. This is displayed in the
upper panel of Fig.~\ref{histogram} which shows that although $\Delta p$
peaks around $0.2$ GeV/c, it can be relatively large in some of the
quark combinations. This is different from the naive analytical
coalescence scenario. Using the original AMPT model, we display the
scaling relation $v_n \sim KE_T$ for initial hadrons right after
hadronization and final hadrons after hadronic rescatterings with a
10 mb parton scattering cross section in Fig.~\ref{vnket_ampt_ori}. One sees
that the flows of initial hadrons do not deviate from the NCQ
scaling relation $v_{n}/n_{q} \sim KE_{T}/n_{q}$ by much, although
$\Delta p$ between valence quark is not small. On the other hand,
the flows of final hadrons follow reasonably well the modified NCQ
scaling $v_{n}/n_{q}^{n/2} \sim KE_{T}/n_{q}$ after hadronic
evolution, consistent with the results in Ref.~\cite{Han11}.

\begin{figure}[h]
\includegraphics[scale=0.8]{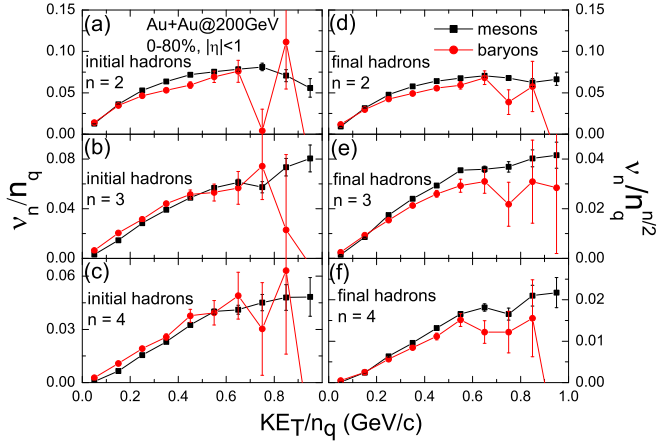} \caption{(Color online)
Same as Fig.~\ref{vnket_ampt_ori} but from the modified AMPT model. } \label{vnket_ampt_mod}
\end{figure}

The results from the original AMPT model didn't tell us whether the
modified NCQ scaling of final hadrons comes from the 'imperfect' coalescence or the
hadronic afterburner effect. To effectively study the hadronic
afterburner effect with a coalescence scenario similar to the
analytical one, we modified the AMPT model by abandoning the hadrons
with $\Delta p$ larger than 0.2 GeV/c, as displayed in the lower
panel of Fig.~\ref{histogram}. The rescatterings and decays of these
hadrons in the hadronic phase are turned off, and they will not
enter the flow analysis by special labelling. In this case the
effective density in the hadronic phase is lower and the hadronic
rescattering effect is weaker. It is seen from Fig.~\ref{vnket_ampt_mod} that
the flows of initial hadrons are closer to the NCQ scaling relation
$v_{n}/n_{q} \sim KE_{T}/n_{q}$ compared with that from the original
AMPT model shown in Fig.~\ref{vnket_ampt_ori}, consistent with the results from
the analytical coalescence scenario except that the magnitude of the
flows is slightly different, as a result of different hadron masses
used in the two approaches. Despite of the weaker hadronic
afterburner effect compared with that from the original AMPT
calculation, the flows of final hadrons again follow the relation
$v_{n}/n_{q}^{n/2} \sim KE_{T}/n_{q}$, after hadronic evolution
including elastic and inelastic scatterings as well as resonance
decays. It is thus more convincible that the hadronic afterburner
can be responsible for the modified NCQ scaling.

\begin{figure}[h]
\includegraphics[scale=0.8]{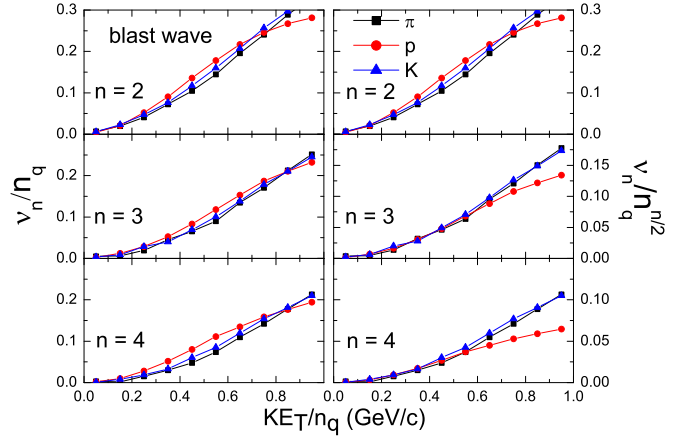} \caption{(Color online)
Scaling relation of $v_{n} \sim KE_{T}$ for pions, kaons, and protons from a generalized blast wave model. }
\label{vnket_blw}
\end{figure}

Since the hadronic rescatterings, which further thermalize the
system, can lead to the modified NCQ scaling relation, one would
expect that the latter might be due to the thermalization mechanism
rather than the coalescence picture. This idea can be tested with
Cooper-Frye freeze-out in the hydrodynamical model or the thermal
blast wave model. Similar analysis has been done in
Ref.~\cite{Pra05}, and in this study we apply a generalized blast
wave model including higher-order flows. The scaling relation of
$v_n \sim KE_T$ is compared in Fig.~\ref{vnket_blw} for pions, kaons, and
protons, and the similar magnitude of $v_n$ for different orders is
due to the same flow parameter used in the generalized blast wave
model as mentioned in Sec.~\ref{bw}. We observed the mass ordering
that the flow of heavy particles is below that of lighter particles
if $v_n$ is plotted as a function of transverse momentum $p_T$.
However, it is seen that flows of pions, kaons, and protons do not
deviate from the NCQ scaling relation $v_{n}/n_{q} \sim
KE_{T}/n_{q}$ by much even from a thermal blast wave model where the
only difference between different particle species is their masses.
On the other hand, it is observed that the modified NCQ scaling
$v_{n}/n_{q}^{n/2} \sim KE_{T}/n_{q}$ is well satisfied for
higher-order anisotropic flows at smaller transverse kinetic
energies. It is of great interest to see whether this is the case in
a more consistent hydrodynamic model and with the hadronic
afterburner effect.

\subsection{The scaling ratio $v_{n}/v_{2}^{n/2}$ }

The scaling relation of $v_4 \sim v_2^2$ has been observed
experimentally~\cite{STAR04a,STAR07,PHENIX10} for
many years, and it has been studied theoretically in both transport
models~\cite{Che04,VPK12} and hydrodynamic
models~\cite{Bor06,Gom10,Gar12}. The general relation of $v_n
\sim v_2^{n/2}$ from consistent event plane analysis was realized
only recently~\cite{ATLAS12}. This scaling relation is important in
understanding the initial condition~\cite{Lac10,Lac11b} and the properties of the
produced QGP~\cite{Lac11a,Gar12}. From the analytical coalescence scenario
in the present study, we will see that the scaling coefficient
depends not only on the viscosity of QGP but on the hadron species
as well.

\begin{figure}[h]
\includegraphics[scale=0.8]{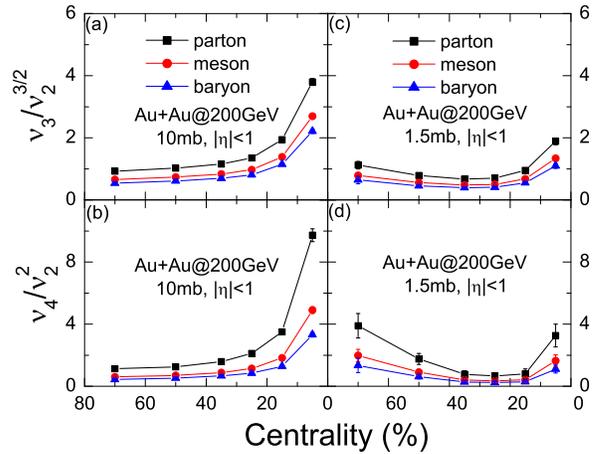}
\caption{(Color online) Centrality dependence of
$v_{n}/v_{2}^{n/2}$ for partons, mesons, and baryons in Au+Au
collisions at $\sqrt{s_{NN}}=200$ GeV from the analytical
coalescence scenario with partonic flows from the parton scattering
cross section of 10 mb (left) and 1.5 mb (right). } \label{vnv2_cen}
\end{figure}

\begin{figure}[h]
\includegraphics[scale=0.8]{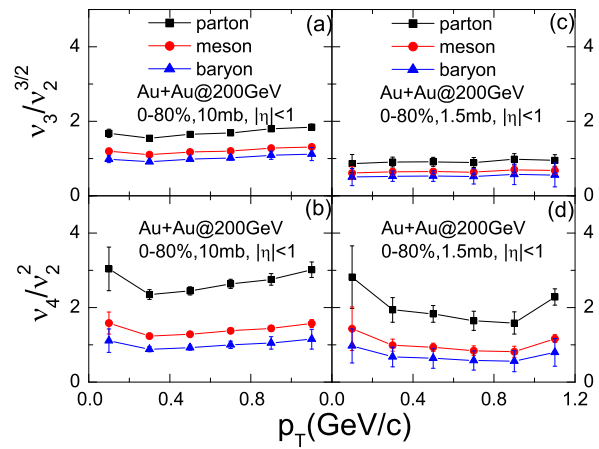}
\caption{(Color online) Transverse momentum ($p_{T}$) dependence of
$v_{n}(p_{T})/[v_{2}(p_{T})]^{n/2}$ in mini-bias Au+Au collisions at
$\sqrt{s_{NN}}=200$ GeV from the analytical coalescence scenario
with partonic flows from the parton scattering cross section of 10
mb (left) and 1.5 mb (right).  } \label{vnv2_pt}
\end{figure}

With the partonic phase-space distribution at freeze-out from the
AMPT model, we have obtained the anisotropic flows for hadrons via
the analytical coalescence scenario and display the scaling relation
of $v_n \sim v_2^{n/2}$ in Figs.~\ref{vnv2_cen} and \ref{vnv2_pt}.
Figure~\ref{vnv2_cen} shows the centrality dependence of the
$v_n/v_2^{n/2}$ ratio for $n=3$ and 4 for partons, mesons, and
baryons, and the results from partonic cross sections of 10 mb and
1.5 mb are compared. Compared with the ratios for partons,
$v_3/v_2^{3/2}$ is about $1/\sqrt{2}$ that for mesons and about
$1/\sqrt{3}$ that for baryons, and $v_4/v_2^2$ is about $1/2$ that
for mesons and about $1/3$ that for baryons, according to
Eqs.~(\ref{r3m}), (\ref{r4m}), (\ref{r3b}), and (\ref{r4b}).
Interestingly, from a parton scattering cross section of 10 mb, the
$v_n/v_2^{n/2}$ ratio decreases with increasing centrality, as a
result of similar centrality dependence of the initial anisotropy
ratio $\epsilon_n/\epsilon_2^{n/2}$ pointed out in
Refs.~\cite{Lac11a,Lac11b}. On the other hand, from a parton
scattering cross section of 1.5 mb, the correlation between the
initial anisotropies $\epsilon_n$ and the final collective flows
$v_n$ is not that strong and the ratio $v_n/v_2^{n/2}$ shows a
non-monotonical dependence on the centrality. The latter case is
similar to that observed from the STAR
Collaboration~\cite{YB07PhD} (also Figs.~4 and 5 in Ref.~\cite{Gom10}). Figures~\ref{vnv2_pt} shows that the
$v_n/v_2^{n/2}$ ratios for partons, mesons, and baryons are mostly
independent of the transverse momentum. This is an interesting
phenomena showing that the QGP interaction generates the anisotropic
flows simultaneously in a $p_T$-independent way according to the
relation $v_n/\epsilon_n \sim (v_2/\epsilon_2)^{n/2}$ from initial
anisotropies $\epsilon_n$~\cite{Lac11a}. It is also
interesting to see that the $v_n/v_2^{n/2}$ ratio is smaller from a
parton scattering cross section of 1.5 mb compared with that of 10
mb, and the effect is larger for $n=3$ than for $n=4$. The insensitivity of $v_4/v_2^2$ to the parton scattering cross section can be due to the strong correlation between $v_2$ and $v_4$. The
$v_n/v_2^{n/2}$ ratio for partons is generated by the initial
condition and the interaction of QGP, while from the analytical
coalescence scenario it is guaranteed that the behavior of mesons
and baryons follow the same centrality and transverse momentum
dependence of that for partons. According to the previous discussion,
the $v_n/v_2^{n/2}$ ratios for mesons and baryons are expected to be
almost the same experimentally, since the modified NCQ scaling
$v_{n}/n_{q}^{n/2} \sim KE_{T}/n_{q}$ is satisfied. To study the
initial condition and the properties of QGP through the
$v_n/v_2^{n/2}$ ratio, the coalescence correction is non-negligible.

In the present work, we have further investigated the effect of
the hadronic afterburner on the $v_n/v_2^{n/2}$ ratio by using the original and modified AMPT model. We found that the effect of hadronic rescattering on $v_n/v_2^{n/2}$ ratio is much smaller compared with that of the partonic interaction. On the other hand, the ratios are similar for mesons and baryons from the AMPT model. This supports our previous discussion on the validity of the modified NCQ scaling law $v_n/n_q^{n/2} \sim KE_T/n_q$, which is approximately satisfied from the AMPT model calculations.


\section{Conclusions}
\label{summary}

In this work, we have investigated the modified
number-of-constituent-quark (NCQ) scaling $v_{n}/n_{q}^{n/2} \sim
KE_{T}/n_{q}$ and the scaling relation $v_{n} \sim v_{2}^{n/2}$. We
found that the modified NCQ scaling can't be obtained from the naive
analytical coalescence scenario, which allows the coalescence of
quarks only if they have the same momentum, even if event-by-event
fluctuations are taken into account. The reason for that is due to
the different scaling coefficients for mesons and baryons in the
scaling relation $v_{n} \sim v_{2}^{n/2}$, while experimentally they
are expected to be almost the same. On the other hand, the modified NCQ
scaling may stem from the hadronic afterburner effect or thermal
freeze-out rather than the coalescence mechanism. The centrality
dependence of the $v_n/v_2^{n/2}$ ratio has been shown to be
sensitive to the parton scattering cross section, while the
$p_T$-independency of the ratio seems to be a robust phenomena. The
$v_3/v_2^{3/2}$ ratio is found to be more sensitive to the partonic
interaction compared with $v_4/v_2^{2}$. Our investigation can be important in understanding
the hadronization mechanism as well as the correlation between the
anisotropic flows and the initial anisotropies of QGP in
relativistic heavy-ion collisions.

\begin{acknowledgments}
We thank Che Ming Ko for helpful comments, Song Zhang for helpful
discussions, and Chen Zhong for maintaining the high-quality
performance of the computer facility. This work was supported by the
Major State Basic Research Development Program (973 Program) of
China under Contract Nos. 2015CB856904 and 2014CB845401, the
National Natural Science Foundation of China under Grant Nos.
11475243 and 11421505, the "100-talent plan" of Shanghai Institute
of Applied Physics under Grant Nos. Y290061011 and Y526011011 from
the Chinese Academy of Sciences, the Shanghai Key Laboratory of
Particle Physics and Cosmology under Grant No. 15DZ2272100, and the
"Shanghai Pujiang Program" under Grant No. 13PJ1410600.
\end{acknowledgments}

\appendices

\renewcommand\thesection{APPENDIX~\Alph{section}}
\renewcommand\theequation{\Alph{section}.\arabic{equation}}
\section{hadronic anisotropic flows from analytical coalescence}
\label{app}

In this appendix, we gave the expressions of the higher-order
harmonic flows of hadrons based on an ideal quark coalescence
scenario by considering the event-by-event initial density
fluctuations. From the azimuthal distribution of partons in terms of
their anisotropic flows up to the fourth order
\begin{eqnarray}
f(p_{T},\phi) &\propto& f_0 + \sum_{n=1}^{4} f_{n} \cos [n(\phi-
\psi_{n})]
\end{eqnarray}%
with $f_0=1$ and $f_n=2v_n$, the azimuthal distribution of mesons in the limit that their valence quarks should have the same momentum can be expressed as
\begin{eqnarray}\label{fm}
&&F(2p_{T},\phi) \propto f^{2}(p_{T},\phi) \notag\\
&&= F_{0}+2f_{0}f_{1}\cos(\phi-\psi_{1}) + f_{1}f_{2}\cos(\phi+\psi_{1}-2\psi_{2})
\notag \\
&&+ f_{2}f_{3}\cos(\phi+2\psi_{2}-3\psi_{3})+ f_{3}f_{4}\cos(\phi+3\psi_{3}-4\psi_{4})
\notag \\
&&+ 2f_{0}f_{2}\cos(2\phi-2\psi_{2})+\frac{1}{2}f_{1}^{2}\cos(2\phi-2\psi_{1})
\notag \\
&&+ f_{1}f_{3}\cos(2\phi+\psi_{1}-3\psi_{3})+f_{2}f_{4}\cos(2\phi+2\psi_{2}-4\psi_{4})
\notag \\
&&+ 2f_{0}f_{3}\cos(3\phi-3\psi_{3})+f_{1}f_{2}\cos(3\phi-\psi_{1}-2\psi_{2})
\notag \\
&&+ f_{1}f_{4}\cos(3\phi+\psi_{1}-4\psi_{4})+ 2f_{0}f_{4}\cos(4\phi-4\psi_{4})
\notag \\
&&+ \frac{1}{2}f_{2}^{2}\cos(4\phi-4\psi_{2})+f_{1}f_{3}\cos(4\phi-\psi_{1}-3\psi_{3})
\notag \\
&&+ ...
\end{eqnarray}%
with
\begin{eqnarray}
F_{0}=f_{0}^{2}+\frac{1}{2}f_{1}^{2}+\frac{1}{2}f_{2}^{2}+\frac{1}{2}f_{3}^{2}+\frac{1}{2}f_{4}^{2}.
\end{eqnarray}%
The anisotropy flows of mesons can be calculated from
\begin{eqnarray}
V_{n} = \frac{\int_{0}^{2\pi} \cos(n\phi - n\psi_{n})F(2p_{T},\phi)d\phi}{\int_{0}^{2\pi}F(2p_{T},\phi)d\phi},
\end{eqnarray}%
and their expressions for different orders are
\begin{eqnarray}
V_{1} &=& \frac{1}{F_{0}}[f_{0}f_{1}+\frac{1}{2}f_{1}f_{2}\cos(2\psi_{1}-2\psi_{2})
\notag \\
&+& \frac{1}{2}f_{2}f_{3}\cos(\psi_{1}+2\psi_{2}-3\psi_{3})
\notag \\
&+& \frac{1}{2}f_{3}f_{4}\cos(\psi_{1}+3\psi_{3}-4\psi_{4})],
\end{eqnarray}%
\begin{eqnarray}
V_{2} &=& \frac{1}{F_{0}}[f_{0}f_{2}+\frac{1}{4}f_{1}^{2}\cos(2\psi_{1}-2\psi_{2})
\notag \\
&+& \frac{1}{2}f_{1}f_{3}\cos(\psi_{1}+2\psi_{2}-3\psi_{3})
\notag \\
&+&\frac{1}{2}f_{2}f_{4}\cos(4\psi_{2}-4\psi_{4})], \label{v2m}
\end{eqnarray}%
\begin{eqnarray}
V_{3} &=& \frac{1}{F_{0}}[f_{0}f_{3}+\frac{1}{2}f_{1}f_{2}\cos(\psi_{1}+2\psi_{2}-3\psi_{3})
\notag \\
&+& \frac{1}{2}f_{1}f_{4}\cos(\psi_{1}+3\psi_{3}-4\psi_{4})], \label{v3m}
\end{eqnarray}%
and
\begin{eqnarray}
V_{4} &=& \frac{1}{F_{0}}[f_{0}f_{4}+ \frac{1}{4}f_{2}^{2}\cos(4\psi_{2}-4\psi_{4})
\notag \\
&+& \frac{1}{2}f_{1}f_{3}\cos(\psi_{1}+3\psi_{3}-4\psi_{4})]. \label{v4m}
\end{eqnarray}%

Similarly, the azimuthal distribution of baryons in the same
scenario can be expressed as
\begin{eqnarray}\label{fb}
&&\widetilde{F}(3p_{T},\phi) \propto f^{3}(p_{T},\phi) \notag \\
&&= \widetilde{F}_{0}+(\frac{3}{4}f_{1}^{3}+3f_{0}^{2}f_{1}+\frac{3}{2}f_{1}f_{2}^{2}
\notag \\
&&+ \frac{3}{2}f_{1}f_{3}^{2}+\frac{3}{2}f_{1}f_{4}^{2})\cos(\phi-\psi_{1})+ 3f_{1}f_{2}\cos(\phi+\psi_{1}-2\psi_{2})
\notag \\
&&+ \frac{3}{4}f_{1}^{2}f_{3}\cos(\phi+2\psi_{1}-3\psi_{3})+ \frac{3}{4}f_{2}^{2}f_{3}\cos(\phi+3\psi_{3}-4\psi_{2})
\notag \\
&&+ 3f_{0}f_{2}f_{3}\cos(\phi+2\psi_{2}-3\psi_{3})+ 3f_{0}f_{3}f_{4}\cos(\phi+3\psi_{3}-4\psi_{4})
\notag \\
&&+ \frac{3}{2}f_{1}f_{2}f_{4}\cos(\phi+\psi_{1}+2\psi_{2}-4\psi_{4})
\notag \\
&&+ \frac{3}{2}f_{2}f_{3}f_{4}\cos(\phi-2\psi_{1}-3\psi_{3}+4\psi_{4})
\notag \\
&&+ (\frac{3}{4}f_{2}^{3}+3f_{0}^{2}f_{2}+\frac{3}{2}f_{2}f_{1}^{2}+ \frac{3}{2}f_{2}f_{3}^{2}+\frac{3}{2}f_{2}f_{4}^{2})\cos(2\phi-2\psi_{2})
\notag \\
&&+ \frac{3}{2}f_{0}f_{1}^{2}\cos(2\phi-2\psi_{1})+ \frac{3}{4}f_{1}^{2}f_{4}\cos(2\phi+2\psi_{1}-4\psi_{4})
\notag \\
&&+ \frac{3}{4}f_{3}^{2}f_{4}\cos(2\phi-6\psi_{3}+4\psi_{4})+ 3f_{0}f_{1}f_{3}\cos(2\phi+\psi_{1}-3\psi_{3})
\notag \\
&&+ 3f_{0}f_{2}f_{4}\cos(2\phi+2\psi_{2}-4\psi_{4})
\notag \\
&&+ \frac{3}{2}f_{1}f_{2}f_{3}\cos(2\phi-\psi_{1}+2\psi_{2}-3\psi_{3})
\notag \\
&&+ \frac{3}{2}f_{1}f_{3}f_{4}\cos(2\phi-\psi_{1}+3\psi_{3}-4\psi_{4})
\notag \\
&&+ (\frac{3}{4}f_{3}^{3}+3f_{0}^{2}f_{3}+\frac{3}{2}f_{3}f_{1}^{2}+ \frac{3}{2}f_{3}f_{2}^{2}+\frac{3}{2}f_{3}f_{4}^{2})\cos(3\phi-3\psi_{3})
\notag \\
&&+ \frac{1}{4}f_{1}^{3}\cos(3\phi-3\psi_{1})+ \frac{3}{4}f_{1}f_{2}^{2}\cos(3\phi+\psi_{1}-4\psi_{2})
\notag \\
&&+ 3f_{1}f_{2}\cos(3\phi-\psi_{1}-2\psi_{2})
+ 3f_{0}f_{1}f_{4}\cos(3\phi+\psi_{1}-4\psi_{4})
\notag \\
&&+ \frac{3}{2}f_{1}f_{2}f_{4}\cos(3\phi-\psi_{1}+2\psi_{2}-4\psi_{4})
\notag \\
&&+ \frac{3}{2}f_{2}f_{3}f_{4}\cos(3\phi-\psi_{2}+3\psi_{3}-4\psi_{4})
\notag \\
&&+ (\frac{3}{4}f_{4}^{3}+3f_{0}^{2}f_{4}+\frac{3}{2}f_{4}f_{1}^{2}
+ \frac{3}{2}f_{4}f_{2}^{2}+\frac{3}{2}f_{4}f_{3}^{2})\cos(4\phi-4\psi_{4})
\notag \\
&&+ \frac{3}{2}f_{0}f_{2}^{2}\cos(4\phi-4\psi_{2})+ \frac{3}{4}f_{1}^{2}f_{1}\cos(4\phi-2\psi_{1}-2\psi_{2})
\notag \\
&&+ 3f_{0}f_{1}f_{3}\cos(4\phi-\psi_{1}-3\psi_{3})
\notag \\
&&+ \frac{3}{2}f_{1}f_{2}f_{3}\cos(4\phi+\psi_{1}-2\psi_{2}-3\psi_{3})
\notag \\
&&+ \frac{3}{4}f_{3}^{2}f_{2}\cos(4\phi+2\psi_{2}-6\psi_{3})
\notag \\
&&+ ...
\end{eqnarray}%
with
\begin{eqnarray}
\widetilde{F}_{0}&=&f_{0}^{3}+\frac{3}{2}f_{0}f_{1}^{2}+\frac{3}{2}f_{0}f_{2}^{2}+\frac{3}{2}f_{0}f_{3}^{2}+\frac{3}{2}f_{0}f_{4}^{2}
\notag \\
&+&\frac{3}{4}f_{1}^{2}f_{2}\cos(2\psi_{1}-2\psi_{2})+\frac{3}{2}f_{1}f_{2}f_{3}\cos(\psi_{1}+2\psi_{2}-3\psi_{3})
\notag \\
&+&\frac{3}{4}f_{2}^{2}f_{4}\cos(4\psi_{2}-4\psi_{4})+\frac{3}{2}f_{1}f_{3}f_{4}\cos(\psi_{1}+3\psi_{3}-4\psi_{4}).\notag\\
\end{eqnarray}

The anisotropic flows of baryons can be calculated from
\begin{eqnarray}
\widetilde{V}_{n} &=& \frac{\int_{0}^{2\pi} \cos(n\phi - n\psi_{n})\widetilde{F}(3p_{T},\phi)d\phi}{\int_{0}^{2\pi}\widetilde{F}(3p_{T},\phi)d\phi},
\end{eqnarray}%
and their detailed expressions are
\begin{eqnarray}
\widetilde{V}_{1} &=& \frac{1}{\widetilde{F}_{0}}[\frac{3}{2}f_{1}f_{0}^{2}+\frac{3}{8}f_{1}^{3}+\frac{3}{4}f_{1}f_{2}^{2}+\frac{3}{4}f_{1}f_{3}^{2}
\notag \\
&+& \frac{3}{4}f_{1}f_{4}^{2}+\frac{3}{2}f_{1}f_{2}\cos(2\psi_{1}-2\psi_{2})
\notag \\
&+& \frac{3}{8}f_{1}^{2}f_{3}\cos(3\psi_{1}-3\psi_{3})
\notag \\
&+& \frac{3}{8}f_{1}^{2}f_{3}\cos(3\psi_{1}-3\psi_{3})
\notag \\
&+& \frac{3}{8}f_{2}^{2}f_{3}\cos(\psi_{1}+3\psi_{3}-4\psi_{2})
\notag \\
&+& \frac{3}{2}f_{0}f_{2}f_{3}\cos(\psi_{1}+2\psi_{2}-3\psi_{3})
\notag \\
&+& \frac{3}{4}f_{1}f_{2}f_{4}\cos(2\psi_{1}+2\psi_{2}-4\psi_{4})
\notag \\
&+& \frac{3}{2}f_{0}f_{3}f_{4}\cos(\psi_{1}+3\psi_{3}-4\psi_{4})
\notag \\
&+& \frac{3}{4}f_{2}f_{3}f_{4}\cos(\psi_{1}-2\psi_{2}-3\psi_{3}+4\psi_{4})],
\end{eqnarray}%
\begin{eqnarray}
\widetilde{V}_{2} &=& \frac{1}{\widetilde{F}_{0}}[\frac{3}{2}f_{2}f_{0}^{2}+\frac{3}{8}f_{2}^{3}+\frac{3}{4}f_{2}f_{1}^{2}+\frac{3}{4}f_{2}f_{3}^{2}
\notag \\
&+& \frac{3}{4}f_{2}f_{4}^{2}+\frac{3}{4}f_{0}f_{1}^{2}\cos(2\psi_{1}-2\psi_{2})
\notag \\
&+& \frac{3}{8}f_{1}^{2}f_{4}\cos(2\psi_{1}+2\psi_{2}-4\psi_{4})
\notag \\
&+& \frac{3}{4}f_{3}^{2}f_{4}\cos(2\psi_{2}+4\psi_{4}-6\psi_{3})
\notag \\
&+& \frac{3}{2}f_{0}f_{1}f_{3}\cos(\psi_{1}+2\psi_{2}-3\psi_{3})
\notag \\
&+& \frac{3}{4}f_{1}f_{2}f_{3}\cos(\psi_{1}+3\psi_{3}-4\psi_{2})
\notag \\
&+& \frac{3}{2}f_{0}f_{2}f_{4}\cos(4\psi_{2}-4\psi_{4})
\notag \\
&+& \frac{3}{4}f_{1}f_{3}f_{4}\cos(\psi_{1}-2\psi_{2}-3\psi_{3}+4\psi_{4})], \label{v2b}
\end{eqnarray}%
\begin{eqnarray}
\widetilde{V}_{3} &=& \frac{1}{\widetilde{F}_{0}}[\frac{3}{2}f_{3}f_{0}^{2}+\frac{3}{8}f_{3}^{3}+\frac{3}{4}f_{3}f_{1}^{2}+\frac{3}{4}f_{3}f_{2}^{2}
\notag \\
&+& \frac{3}{4}f_{3}f_{4}^{2}+\frac{3}{8}f_{1}^{3}\cos(3\psi_{1}-3\psi_{3})
\notag \\
&+& \frac{3}{8}f_{2}^{2}f_{1}\cos(\psi_{1}+3\psi_{3}-4\psi_{2})
\notag \\
&+& \frac{3}{2}f_{1}f_{2}\cos(\psi_{1}+2\psi_{2}-3\psi_{3})
\notag \\
&+& \frac{3}{2}f_{0}f_{1}f_{4}\cos(\psi_{1}+3\psi_{3}-4\psi_{4})
\notag \\
&+& \frac{3}{4}f_{2}f_{3}f_{4}\cos(2\psi_{2}+4\psi_{4}-6\psi_{3})
\notag \\
&+& \frac{3}{4}f_{1}f_{2}f_{4}\cos(\psi-2\psi_{2}-3\psi_{3}+4\psi_{4})], \label{v3b}
\end{eqnarray}%
and
\begin{eqnarray}
\widetilde{V}_{4} &=& \frac{1}{\widetilde{F}_{0}}[\frac{3}{2}f_{4}f_{0}^{2}+\frac{3}{8}f_{4}^{3}+\frac{3}{4}f_{4}f_{1}^{2}+\frac{3}{4}f_{4}f_{2}^{2}
\notag \\
&+& \frac{3}{4}f_{4}f_{3}^{2}+\frac{3}{4}f_{0}f_{2}^{2}\cos(4\psi_{2}-4\psi_{4})
\notag \\
&+& \frac{3}{8}f_{1}^{2}f_{2}\cos(2\psi_{1}+2\psi_{2}-4\psi_{4})
\notag \\
&+& \frac{3}{2}f_{0}f_{1}f_{3}\cos(\psi_{1}+3\psi_{3}-4\psi_{4})
\notag \\
&+& \frac{3}{4}f_{1}f_{2}f_{3}\cos(\psi_{1}-2\psi_{2}-3\psi_{3}+4\psi_{4})
\notag \\
&+& \frac{3}{8}f_{2}f_{3}^{2}\cos(2\psi_{2}+4\psi_{4}-6\psi_{3})]. \label{v4b}
\end{eqnarray}%
Here we only consider the flows up to the fourth order, thus the even higher-order terms in Eqs.~(\ref{fm}) and (\ref{fb}) do not contribute.

To investigate the scaling relation between flows of different
orders $v_n \sim v_2^{n/2}$, we also give the expressions of the
corresponding ratios in the analytical coalescence scenario. The
ratios of $V_n/V_2^{n/2}$ for mesons with $n=3$ and 4 by neglecting
the higher-order terms can be written as
\begin{eqnarray}
\frac{V_{3}}{V_{2}^{3/2}} &\approx& F_{0}^{1/2}[\frac{1}{\sqrt2}\frac{v _{3}}{v _{2}^{3/2}}
+ \frac{1}{\sqrt2}\frac{v_{1}}{v_{2}^{1/2}} \cos(\psi _{1} + 2\psi_{2} - 3\psi_{3})
\notag \\
&+& \frac{1}{\sqrt2}\frac{v _{1}v _{4}}{v _{2}^{3/2}}\cos(\psi_{1}+3\psi_{3}-4\psi_{4})], \label{r3m} \\
\frac{V_{4}}{V_{2}^{2}} &\approx& F_{0}[\frac{1}{2}\frac{v _{4}}{v _{2}^{2}}+\frac{1}{4}\cos(4\psi_{2}-4\psi_{4})
\notag \\
&+& \frac{1}{2}\frac{v _{1}v _{3}}{v _{2}^{2}}\cos(\psi_{1}+3\psi_{3}-4\psi_{4})]. \label{r4m}
\end{eqnarray}%
The ratios of  $\widetilde{V}_{n}/\widetilde{V}_2^{n/2}$ for baryons
with $n=3$ and 4 by neglecting the higher-order terms can be written
as
\begin{eqnarray}
\frac{\widetilde{V}_{3}}{\widetilde{V}_{2}^{3/2}} &\approx& \widetilde{F}_{0}^{1/2}[\frac{1}{\sqrt3}\frac{v _{3}}{v _{2}^{3/2}}
+\frac{1}{\sqrt3}\frac{v _{3}^{3}}{v _{2}^{3/2}}
+ \frac{2}{\sqrt3}\frac{v _{3}v_{1}^{2}}{v_{2}^{3/2}}
\notag \\
&+& \frac{2}{\sqrt3}v_{3}v_{2}^{1/2} + \frac{2}{\sqrt3}\frac{v _{3}v_{4}^{2}}{v _{2}^{3/2}}
\notag \\
&+& \frac{1}{\sqrt3}\frac{v _{1}^{3}}{v _{2}^{3/2}}\cos(3\psi_{1}-3\psi_{3})
\notag \\
&+& \frac{1}{\sqrt3}v _{2}^{1/2}v _{1}\cos(\psi_{1}+3\psi_{3}-4\psi_{2})
\notag \\
&+& \frac{2}{\sqrt3}\frac{v _{1}}{v _{2}^{1/2}}\cos(\psi_{1}+2\psi_{2}-3\psi_{3})
\notag \\
&+& \frac{2}{\sqrt3}\frac{v _{1}v_{4}}{v _{2}^{3/2}}\cos(\psi_{1}+3\psi_{3}-4\psi_{4})
\notag \\
&+& \frac{2}{\sqrt3}\frac{v _{3}v_{4}}{v _{2}^{1/2}}\cos(2\psi_{2}+4\psi_{4}-6\psi_{3})
\notag \\
&+& \frac{2}{\sqrt3}\frac{v _{1}v_{4}}{v _{2}^{1/2}}\cos(\psi_{1}-2\psi_{2}-3\psi_{3}+4\psi_{4})], \label{r3b} \notag\\
\\
\frac{\widetilde{V}_{4}}{\widetilde{V}_{2}^{2}} &\approx& \widetilde{F}_{0}[\frac{1}{3}\frac{v _{4}}{v _{2}^{2}}
+ \frac{1}{3}\frac{v _{4}^{3}}{v _{2}^{2}}+\frac{2}{3}\frac{v _{4}v_{1}^{2}}{v _{2}^{2}}
+ \frac{2}{3}v_{4}
\notag \\
&+& \frac{2}{3}\frac{v _{4}v_{3}^{2}}{v _{2}^{2}}+\frac{1}{3}\cos(4\psi_{2}-4\psi_{4})
\notag \\
&+& \frac{1}{3}\frac{v _{1}^{2}}{v _{2}}\cos(2\psi_{1}+2\psi_{2}-4\psi_{4})
\notag \\
&+& \frac{2}{3}\frac{v _{1}v_{3}}{v _{2}^{2}}\cos(\psi_{1}+3\psi_{3}-4\psi_{4})
\notag \\
&+& \frac{1}{3}\frac{v _{3}^{2}}{v _{2}}\cos(2\psi_{2}+4\psi_{4}-6\psi_{3})
\notag \\
&+& \frac{2}{3}\frac{v _{1}v_{3}}{v _{2}}\cos(\psi_{1}-2\psi_{2}-3\psi_{3}+4\psi_{4})]. \label{r4b}
\end{eqnarray}%

\renewcommand\thesection{APPENDIX~\Alph{section}}
\renewcommand\theequation{\Alph{section}.\arabic{equation}}
\section{anisotropic flows from event plane method}
\label{vn}

Here we briefly review the standard method of calculating the
anisotropic flows as well as the event plane from particle
freeze-out distribution in the present work. We refer the readers to
Refs.~\cite{Pos98,Vol08} for more details.

We start from the momentum distribution of emitted particles as
follows:
\begin{eqnarray}
E\frac{d^{3}N}{d^{3}p} &=& \frac{1}{2\pi} \frac{d^{2}N}{p_{T}dp_{T}dy}\left\{1+\sum_{n=1}^{\infty}2v_{n}\cos[n(\phi
-\psi_{n})]\right\}, \notag\\
\end{eqnarray}%
where $\phi$ is the azimuthal angle of emitted particles, $y$ and
$p_T$ are respectively the rapidity and transverse momentum, $v_n$ is the
$n$th-order anisotropic flows, and $\psi_{n}$ is the corresponding
event plane angle. The relation between the event flow vector
$Q_{n}$ and the event plane angle $\psi_{n}$ can be expressed as
\begin{eqnarray}
Q_{n,x} = Q_{n}\cos(n\psi_{n}) = \sum_{i} \omega_{i}\cos(n\phi_{i}), \\
Q_{n,y} = Q_{n}\sin(n\psi_{n}) = \sum_{i} \omega_{i}\sin(n\phi_{i}),
\end{eqnarray}%
where the summation goes over all particles $i$ used in the event
plane calculation, and $\phi_{i}$ and $\omega_{i}$ are respectively
the azimuthal angle and the weight factor for particle $i$, with the
latter set as the transverse momentum of the particle. The event
plane angle can thus be calculated from
\begin{eqnarray}
\psi_{n}=\left[\text{atan2} \frac{\sum_{i} \omega_{i} \sin(n\phi_{i})}{\sum_{i} \omega_{i} \cos(n\phi_{i})}\right]/n.
\end{eqnarray}%
The $n$th-order flow magnitude $v_{n}^{obs}$  with respect to this
event plane is
\begin{eqnarray}
v_{n}^{obs}(p_{T},y) = \langle \cos[n(\phi_{i}-\psi_{n})]\rangle,
\end{eqnarray}%
where $\langle ... \rangle$ denotes an average over all particles in
all events with their azimuthal angle $\phi_{i}$ for a given
rapidity $y$ and transverse momentum $p_{T}$. To remove
auto-correlations, one has to subtract the contribution of the
particle of interest from the total $Q_n$ vector, obtaining a $\psi_{n}$
uncorrelated with that particle. Since finite multiplicity limits
the estimation of the event plane angle, $v_{n}$ has to be corrected
by the event plane resolution for each $n$ given by
\begin{eqnarray}
\label{resolu}
\Re_{n}(\chi) &=& \frac{\sqrt{\pi}}{2} \chi \exp(-\chi^{2}/2)[I_{(k-1)/2}(\chi^{2}/2)
\notag \\
&+& I_{(k+1)/2}(\chi^{2}/2)],
\end{eqnarray}%
where we have $\chi=v_{n}\sqrt{M}$ with $M$ being the particle
multiplicity, and $I_{k}$ is the modified Bessel function. To
calculate the event plane resolution, the full events are divided up
into two independent sub-events of equal multiplicity. Thus the
resolution for sub-events is just the square-root of this
correlation defined as
\begin{eqnarray}
\label{sub}
\Re_{n}^{sub} = \sqrt{ \langle \cos[n(\psi_{n}^{A}-\psi_{n}^{B})] \rangle },
\end{eqnarray}%
where A and B denote the two subgroups of particles. In our
calculation we divided particles within pseudorapidity window
$|\eta|<1$ into two groups of forward and backward spheres with a
gap of $|\Delta\eta|<0.1$. The full event plane resolution is
obtained by
\begin{eqnarray}
\Re_n^{full} = \Re(\sqrt{2} \chi_{sub}),
\end{eqnarray}%
where $\chi_{sub}$ is inversely obtained from the sub-event
resolution $\Re_{n}^{sub}$ via Eq.~(\ref{resolu}). The final
anisotropic flow is
\begin{eqnarray}
v_{n} = \frac{v_{n}^{obs}(p_{T},y)}{\Re_{n}^{full}}.
\end{eqnarray}%

\end{document}